\documentclass[times,twocolumn,final]{elsarticle}

\usepackage{cnf}

\usepackage{framed,multirow}

\usepackage{amssymb}
\usepackage{latexsym}

\usepackage{url}
\usepackage{xcolor}
\definecolor{newcolor}{rgb}{.8,.349,.1}

\usepackage{amsmath}
\usepackage{graphicx}
\usepackage{lineno}
\usepackage{hyperref}

\journal{ }

\begin{document}

\verso{Nie et al.}

\begin{frontmatter}

\title{Pulsation of Burner-Stabilized CH$_4$-O$_2$ Flames Moderated by CO$_2$ Addition}

\author[1]{Xiangyu \snm{Nie}}
\author[1]{Shuoxun \snm{Zhang}}
\author[1]{Shengkai \snm{Wang}\corref{cor1}}

\cortext[cor1]{Corresponding author}
\emailauthor{sk.wang@pku.edu.cn}{Shengkai Wang}

\address[1]{SKLTCS, CAPT, School of Mechanics and Engineering Science, Peking University, 5 Yiheyuan Road, Beijing, 100871, China}

\begin{abstract}
This study investigated the pulsating instability of burner-stabilized premixed CH$_4$-O$_2$ flames at various levels of CO$_2$ dilution with non-zero heat loss. Experiments were conducted using a water-cooled porous-plug burner of 18 mm diameter over a wide range of mixture compositions and flow rates, during which time-resolved measurements of flame chemiluminescence and gas temperature were obtained. The primary oscillation frequencies of the pulsating flames were determined using fast Fourier transform and harmonic power analysis. Phase-locked analysis of the chemiluminescence images revealed an interesting mode-transition phenomenon in the flame oscillations. Under fuel-rich conditions with relatively low heat release rates and low flow rates, the flames exhibited quasi-periodic single-mode oscillations. At elevated flow rates, these oscillations were modulated by low-frequency flame flickering instabilities, which created sidebands around the primary oscillation frequency. At higher heat release rates, the flickering instability further triggered mode splitting, eventually leading to multi-mode oscillations. Regime diagrams of the flame oscillation modes, as well as the stability boundaries, were obtained under various fuel flow rates. These findings can be useful for both fundamental research on flame dynamics and practical applications of CO$_2$-moderated oxy-combustion.
\end{abstract}


\end{frontmatter}

\section*{Novelty and significance statement}
This work presents, to the authors' knowledge, the first systematic measurements of pulsating flame instabilities in burner-stabilized premixed methane–oxygen flames with CO$_2$ addition and wall heat loss. Compared with freely propagating flames, where thermo-diffusive pulsations only occur at high Lewis numbers and Zel'dovich numbers, such pulsations are observed over a much wider range of conditions in the present study. Under normal gravity, the pulsating flames are affected by buoyancy-induced low-frequency flickering, which modulates the pulsation amplitude at low thermal power and can further trigger mode splitting at elevated thermal power. A data set comprising regime diagrams of pulsation mode, the primary frequencies of pulsating flames, and the spatiotemporal evolution of flame chemiluminescence and gas temperature, is established. The results of the present study provide new physical insights into intrinsic flame instabilities and promise to aid practical applications of CO$_2$-moderated oxy-combustion as well.

\section{Introduction}
\label{sec1}
Combustion instabilities, caused either by the interaction of reacting flow with combustor components or by intrinsic processes related to flame dynamics, pose critical challenges to the modern design of power generation and propulsion systems aimed at achieving cleaner and more efficient operations at higher safety levels \cite{candel2002combustion, poinsot2017prediction, lieuwen2021unsteady}, especially amid the increasing adoption of renewable fuels \cite{pitsch2024transition} and carbon-neutral combustion strategies \cite{abubakar2021review}. For example, fuels obtained from renewable sources typically contain a significant portion of lightweight molecules that exhibit kinetic, thermodynamic, and transport properties drastically different from those of conventional large hydrocarbon fossil fuels. These properties can lead to intrinsic flame instabilities within the combustion process itself, even in the absence of surrounding components \cite{matalon2007intrinsic}.

An important type of such instability is the thermo–diffusive instability, which arises from the imbalance between heat and mass transfer \cite{sivashinsky1983instabilities, buckmaster1993structure, clavin1994premixed}. Depending on the effective Lewis number ($Le$) of the reactive mixture, the flame can exhibit either spatial cellular structures (at $Le < 1$) or temporal pulsation of the reaction front (at $Le > 1$). Although the former has been extensively investigated, especially in lean combustion systems of lightweight fuels such as hydrogen and methane \cite{law2005cellular, kadowaki2005unstable, yu2013onset, jin2015cellular, lapalme2018characterization, kim2020laminar, jiang2020cellular, antar2023experimental, zirwes2024role}, experimental studies on the latter remain relatively scarce. 

From a physics perspective, this scarcity is likely due to the onset criterion for pulsating instabilities, which renders experimental observation difficult. According to the classic theoretical analysis of Sivashinsky \cite{sivashinsky1977diffusional}, pulsating instability in a freely propagating premixed adiabatic flame can occur only when $Ze(Le-1) > 4(1+\sqrt{3})$, where $Ze$ is the Zel'dovich number representing the non-dimensional overall activation energy of the mixture. This requires the Lewis number to be impractically high for most fuel-air mixtures \cite{mislavskii2021diffusive}, except for very rich hydrogen/air mixtures \cite{gubernov2017hydrogen}. Though not being an issue for theoretical and computational studies \cite{christiansen2001steady, sung2002stretch}, where arbitrarily large Lewis numbers can be accessed \cite{yuan2006pulsating, wu2012asymptotic, cai2020effects}, it has posed a substantial challenge for general experimental measurements of pulsating instability in freely propagating flames.

In light of this challenge, researchers have resorted to alternative platforms, such as the porous-plug burner, for the experimental investigation of flame pulsation dynamics (for example, see \cite{kurdyumov2008porous, kurdyumov2013influence}). This type of burner features a flat surface made of porous metal plug, which generates a uniform velocity profile and also acts as a heat sink that helps to anchor the flame. The presence of a porous plug amplifies the flame instability through interactions with the burner's boundary conditions, thereby making the pulsation phenomena observable over a wider range of conditions \cite{margolis1981effects}, especially at much lower Lewis numbers that are more relevant to practical engines and combustors. Furthermore, from a pragmatic perspective, this configuration perhaps better represents realistic situations, as all practical combustors have boundaries.

Some pioneering investigations of thermo-diffusive pulsation in burner-stabilized flames date back to the 1980s and 1990s. For example, Blackshear et al. \cite{blackshear1984experimental} observed flame front oscillations in lean propane and rich methane–air mixtures, and Pearlman and Ronney \cite{pearlman1997target} studied a variety of pulsation modes (including concentric rings and rotating spiral waves) in methane, propane, and butane flames in air. Recently, there has been a proliferation of experimental and computational studies in this area (for example, see \cite{gubernov2017hydrogen, nechipurenko2020experimental, mislavskii2021diffusive, moroshkina2023burner, moroshkina2024performance, volkov2024relaxational, li2025two}), focusing primarily on methane-air and hydrogen-air flames. However, for flames with enriched or pure oxygen as the oxidizer, a comprehensive characterization of the pulsating instability remains lacking. These flames are expected to exhibit substantially different instability boundaries and oscillation behaviors due to their higher volumetric heat release rates.

Extended study on this topic is important not only for advancing fundamental flame dynamics research but also for practical applications of oxy-fuel combustion. One prominent application is in rocket engines powered by liquid oxygen and hydrocarbon fuels \cite{neill2009practical}. The successful operation of advanced rocket thrust chambers demands robust mitigation of combustion instabilities, particularly inside the gas generator, where combustion of the methane-oxygen mixture occurs at fuel-rich conditions and can be subject to thermo-diffusive pulsation \cite{boulal2022flame}. Aside from propulsion applications, oxy-combustion is also receiving increased attention in power generation as a promising decarbonization strategy that simplifies CO$_2$ capture by eliminating nitrogen from the exhaust stream \cite{singh2003techno, stanger2015oxyfuel}. In such systems, flue gas recirculation is often required to moderate flame temperature \cite{andersson2007flame, heil2011experimental}, making it essential to account for the effects of CO$_2$ dilution as well. However, as previous studies have suggested \cite{abubakar2021review, kobayashi2007effects}, the flame instability characteristics of O$_2$/CO$_2$ combustion systems are quite different from combustion in air, because (a) the oxygen enrichment modifies the Zel'dovich number by changing the effective active energy and raising the adiabatic flame temperature, and (b) the introduction of CO$_2$ molecules in the unburnt gas mixture modifies its effective Lewis number. These instability characteristics merit further investigation.

The current study aims to extend previous knowledge of pulsating flame instability by systematically exploring the stability boundary, dominant frequency, and modal characteristics of this instability under conditions of oxy-combustion and more realistic thermal boundaries with heat loss. The rest of this paper is organized as follows: Section 2 describes the experimental methods and data analysis procedures; Section 3 presents the measurement results and discusses the frequencies, modes, and regime diagrams of the observed flame oscillation phenomena; and Section 4 offers concluding remarks.

\section{Methods}
\subsection{Experimental method}
\begin{figure*}[h!]
\centering
\includegraphics[width= 0.9\linewidth]{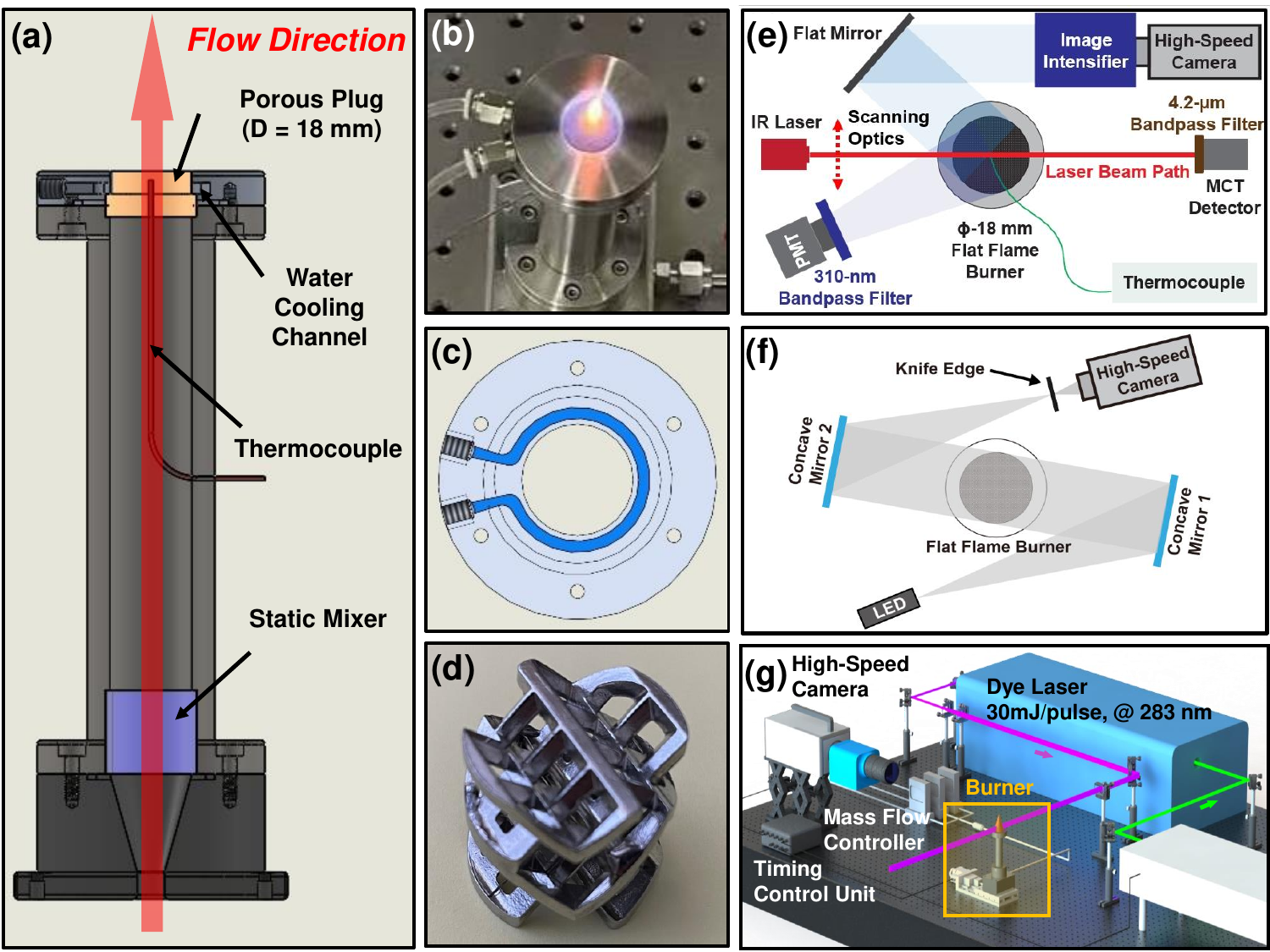}
\caption{\footnotesize The current experimental setup. (a) Configuration of the porous-plug burner. (b) Example of a burner-stabilized flame. (c) Detailed view of the water cooling channel. (d) Photograph of the static mixer. (e) Schematic of the flame chemiluminescence and tunable diode laser diagnostics. (f) The schlieren diagnostic system. (g) The OH-PLIF diagnostic system.}
\label{Fig_01}
\end{figure*}

The current flame experiments were conducted on a custom-built McKenna-type burner, as shown in Fig. \ref{Fig_01}. The burner featured a circular sintered bronze plug of 18 mm diameter that was water-cooled to maintain a stable temperature. The temperature of the porous plug was continuously monitored using an embedded K-type thermocouple (Omega TJ36-CAXL-020-12). High-purity CH$_4$ (99.99\%-grade), O$_2$ (99.999\%-grade), and CO$_2$ (99.999\%-grade) were supplied to the bottom of the burner, where they were thoroughly mixed by an in-line static mixer before passing through the porous plug. The flow rates of fuel, $\rm O_2$, and $\rm CO_2$ were precisely controlled by three Alicat MC series mass flow controllers, with typical uncertainties of 0.1\%, 0.2\%, and 0.5\%, respectively.

Multiple diagnostic methods were employed to investigate the flame oscillation dynamics, including: (1) a spatially and spectrally modulated multi-color laser absorption system for gas temperature measurements (10 kHz, 2\% accuracy); (2) an image-intensified high-speed camera for spatiotemporally resolved imaging of the line-of-sight integrated OH* chemiluminescence (10 kHz, used as a proxy for the local heat release rate); (3) a photomultiplier tube (PMT) for microsecond-resolved measurements of global chemiluminescence variations and transient phenomena; (4) a Z-type high-speed schlieren system; and (5) a pulsed OH-PLIF diagnostic for side-view measurements of the flames. 

The current laser absorption diagnostic used two distributed-feedback interband-cascade lasers (Nanoplus) to probe 12 absorption features of hot CO$_2$ molecules near the $\nu_3$ bandhead at a scan rate of 10 kHz. These transitions exhibit markedly different temperature sensitivities. From the relative intensities of these transitions, the gas temperatures were determined with high sensitivity and robustness. Spatially resolved measurements were enabled by a 2D high-speed parallel beam scanning system consisting of a 2D galvo scanner (Thorlabs, GVS102 2D) and a pair of off-axis parabolic mirrors placed upstream and downstream of the measurement region. The galvo scanner was positioned at the focus of the upstream off-axis parabolic mirror, ensuring that the reflected beams remained parallel to its axis. The downstream mirror focused the parallel beams onto a photodetector that was bandpass-filtered to block the background infrared emission of the flames. A planar measurement speed of 200 Hz was routinely achieved. Further details of this diagnostic method have been documented in a separate work of the authors \cite{zhang2026} and are omitted here for brevity.

Time-synchronized measurements of flame chemiluminescence were conducted using an image-intensified high-speed monochromatic camera (Phantom v611) at a frame rate of 10 kHz and a bit depth of 12 bits. To avoid saturation effects, the gain of the image intensifier (Intelligent Scientific Systems, Model EyeiTS) was adjusted to yield a maximum pixel value of approximately 75\% of the camera's full range. For top-view measurements of the flames, a flat mirror was placed 20 cm above the burner. This imaging setup provided a depth of field of approximately 6 mm, effectively restricting the recorded emission to a narrow region surrounding the flame front while blurring the out-of-focus signal from further downstream. This enabled selective visualization of the reaction zone and minimized interference from background emission. Further details of the imaging setup were documented in a previous study by the authors \cite{wang2025self}. The flames were recorded at a pixel resolution of 256 x 256 and a physical resolution of 0.11 mm x 0.11 mm per pixel. During each measurement, a sequence of 10,000 consecutive images was recorded.

Side-view measurements of the total OH* chemiluminescence were conducted using a Hamamatsu H3695-10 PMT located approximately 6 cm from the burner. A bandpass filter (center wavelength = 310 nm, FWHM = 10 nm) was placed in front of the PMT to reject interfering emissions. The current output of the PMT was converted to a voltage signal by a 10-k$\Omega$ resistor. The -3 dB bandwidth of the PMT measurements, as determined by the transimpedance gain (10-k$\Omega$) and the parasitic capacitance (160 pF) of the signal line, was approximately 100 kHz, corresponding to an effective temporal resolution of $\tau_{1/e} = \rm 1.6 \ \mu s$. The PMT signal of OH* chemiluminescence was recorded simultaneously with the MCT detector signal of transmitted laser intensity at 100 MS/s using a two-channel data acquisition module (National Instruments, PXI-5122).

Side-view schlieren imaging of the flames was conducted with a Z-type configuration using a pair of concave mirrors, a 5-W LED, and a vertical knife edge positioned at the focal plane of the downstream mirror. The knife edge was slightly offset from the focus to enhance image contrast, enabling clear observation of minute density variations in the flow field during the experiment. Schlieren images were recorded with the Phantom v611 high-speed camera at a frame rate of 1 kHz.

OH-PLIF measurements were performed using a nanosecond-pulsed tunable dye laser (LiOPStar-N, LIOP-TEC) at a repetition rate of 10 Hz. The laser wavelength was tuned to 282.997 nm, targeting the P(1.5) + Q(1.5) + R(2.5) transition cluster of the OH A–X (1,0) band, with a pulse energy of approximately 30 mJ. The laser beam was expanded into a horizontal light sheet using a set of cylindrical lenses, achieving a $2\sigma$ thickness of less than 0.3 mm across the burner surface. The OH-PLIF signal was collected with the same intensified camera used for OH* chemiluminescence, spectrally filtered between 300–320 nm to suppress background flame emission. The camera was gated (with an exposure time of 10 $\mu s$) and synchronized to the laser pulses. During each OH-PLIF measurement, the PMT signal of OH* chemiluminescence was recorded simultaneously. The superposition of laser-induced fluorescence and laser scattering on OH* chemiluminescence produced local spikes in the PMT signal, enabling determination of the relative phase of each OH-PLIF image within the flame oscillation cycle.

\subsection{Frequency analysis and phase-locked cycle averaging of the chemiluminescence signal \label{sec:freq_sweep}}

The frequency content of the total chemiluminescence signal was first analyzed using fast Fourier transform (FFT). The results were expressed in terms of normalized power spectral density (NPSD), as shown in Eqn. (1).

\begin{equation} \label{Eqn1}
S(f_k) = \frac{1}{N \Delta f} \cdot \frac{|\sum_{i=1}^{N} I(t_i) e^{-2 \pi j f_k t_i} |^2}{\sum_{i=1}^{N} I(t_i)^2}
\end{equation}

In Eqn. (1), $f_k = k \Delta f$ denotes the k-th discrete Fourier frequency, with $\Delta f$ being the reciprocal of the total measurement time $T = N \Delta t$ (1 s in the current study). $S(f_k)$ denotes the NPSD of the chemiluminescence signal evaluated at $f_k$. $I(t_i)$ represents the discrete-time chemiluminescence signal sampled at $t_i = i \Delta t$, and $j$ denotes the imaginary unit. $S(f_k)$ is also expressed in logarithmic units (dB/Hz) as follows.

\begin{equation} \label{Eqn2}
\hat{S}(f_k) = 10 \log_{10} S(f_k)
\end{equation}

The fundamental frequency of oscillation, $f_{osc}$, was extracted using a harmonic power analysis method developed in a previous study of the authors \cite{wang2025self}. Specifically, $f_{osc}$ was determined by solving a constrained optimization problem, namely maximizing the total power of harmonics within a search window $[f_{min}, f_{max}]$ estimated from the NPSD result, as shown below.

\begin{equation} \label{Eqn3}
f_{osc}=\underset{[f_{min}, f_{max}]}{\operatorname{argmax}}\sum_{k=1}^M\left|\sum_{i=1}^N I(t_i) e^{-2 \pi j k f t_i}\right|^2
\end{equation}

This optimization problem was solved numerically using a brute-force search algorithm, and to improve computational efficiency and robustness, the number of harmonics was truncated to M = 5. In the current study, this method was able to determine the primary oscillation frequency with a typical uncertainty of less than 0.1 Hz (defined by the frequency offset at 95\% maximum total power). Once the primary oscillation frequency was determined, the chemiluminescence images of the pulsating flames were further analyzed using phase-locked cycle averaging, following the same procedure as explained in \cite{wang2025self}. 

\subsection{Axisymmetric reconstruction of temperature fields from laser absorption measurement}

\begin{figure*}[h!]
    \centering
    \includegraphics[width= 0.9\linewidth]{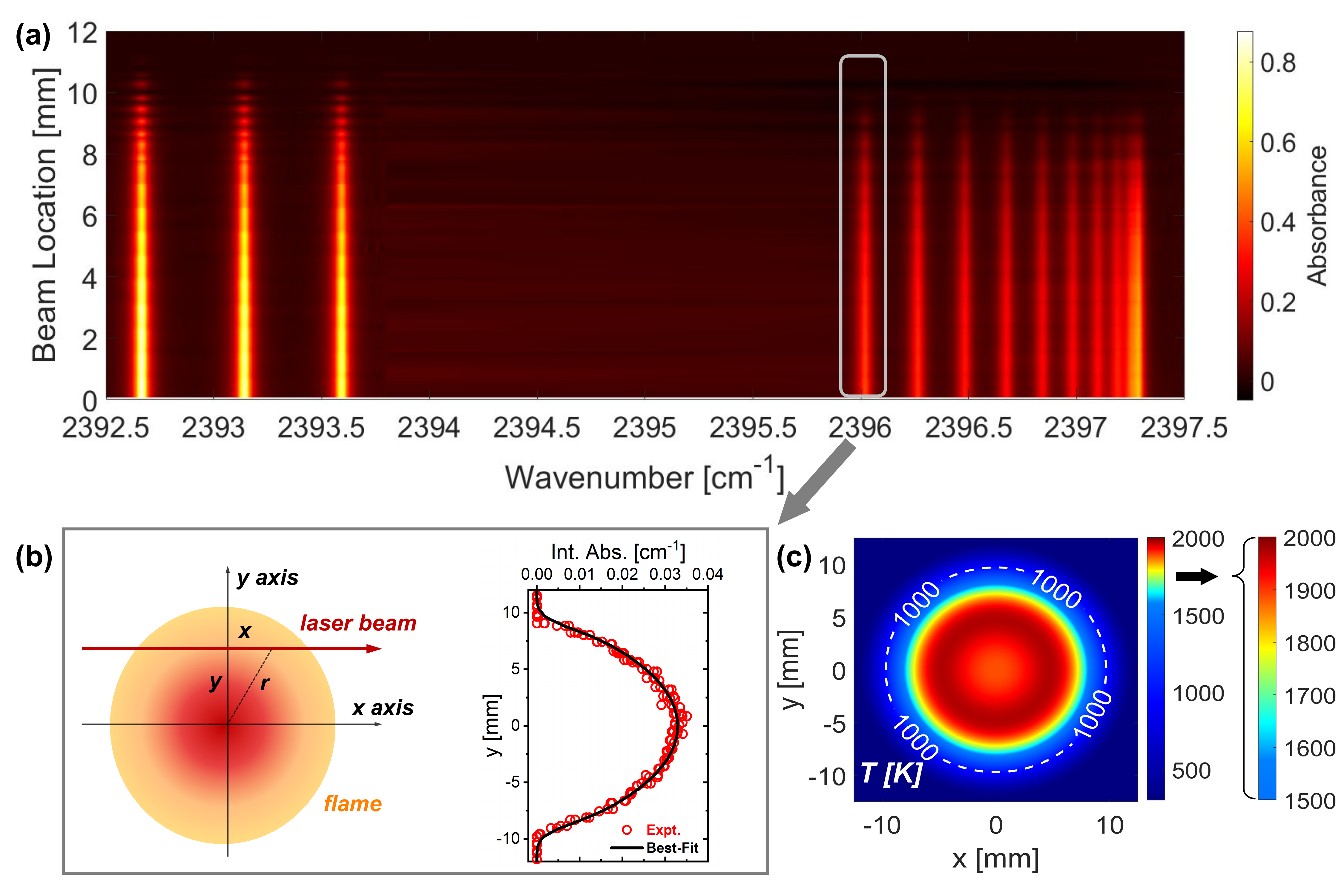}
    \caption{\footnotesize Example measurement of a premixed CH$_4$-O$_2$ flame pulsating in a single axisymmetric mode under the conditions of $\dot{m}_{\rm CH_{4}}$ = 0.20 SLPM and $\dot{m}_{\rm O_{2}}$ = 0.30 SLPM. (a) A spectrogram of the absorbance signal at different beam locations, with data obtained from multiple oscillation cycles at a fixed relative time of 10 ms. (b) The integrated absorbance of an isolated feature at different beam locations. (c) The spatial distribution of gas temperature reconstructed from the integrated absorbance under axial symmetry assumption.}
    \label{Fig_02}
\end{figure*}

Spatially resolved measurements of local gas temperature were achieved by scanning the laser beams across the flame. An example measurement of a pulsating CH$_4$-O$_2$ flame at lower flow rates ($\dot{m}_{\rm CH_{4}}$ = 0.20 SLPM and $\dot{m}_{\rm O_{2}}$ = 0.30 SLPM, respectively) is illustrated in Fig. \ref{Fig_02}. Under these conditions, the flame exhibited a regular mode of oscillation that was periodic in time and nearly axisymmetric in space, as confirmed by time-synchronized chemiluminescence imaging. For oscillating flames with axial symmetry, a reconstruction scheme based on constrained optimization was developed to infer the spatial distribution of temperature from the absorbance data at different beam locations. Further details of the reconstruction scheme are provided in a previous study by the authors \cite{zhang2026} and are omitted here for brevity.

\section{Results and Discussion \label{sec:results}} \addvspace{10pt}
\addvspace{10pt}

A total of 838 pulsating flame experiments were conducted over equivalence ratios ($\phi$) from 1.18 to 2.35 and fuel fractions ($\eta$, defined as $\dot{m}_{CH_4} / (\dot{m}_{CH_4} + \dot{m}_{CO_2})$) from 0.26 to 1.00. The primary oscillation frequencies, oscillation modes, regime diagrams, and fuel dependence of these pulsating flames are discussed below. A summary of the measured frequencies and modes is provided in Supplementary Material I.

\subsection{Pulsation frequencies \label{sec:spectra}}
\addvspace{10pt}
The primary oscillation frequency ($f_{osc}$) of a pulsating flame was determined using the method described in Section 2.2. Across the range of the current experimental conditions, this frequency varied from 26.4 to 1866.0 Hz, exhibiting a negative dependence on the equivalence ratio and a positive dependence on the fuel fraction. 

\begin{figure*}[h!]
    \centering
    \includegraphics[width=0.9\linewidth]{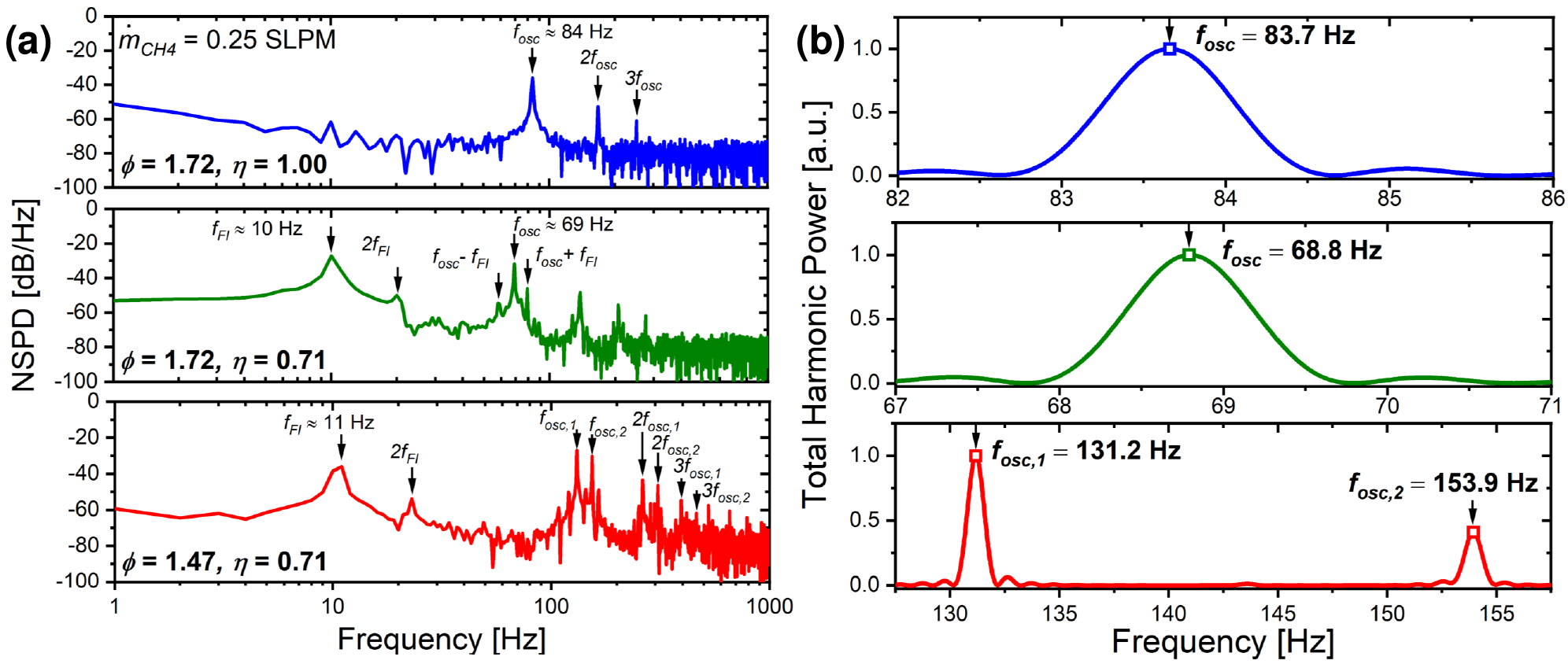}
    \caption{(a) Representative OH* NPSD spectra of pulsating flames at three conditions. Top (blue): $\phi$ = 1.72, $\eta$ = 1.00; middle (green): $\phi$ = 1.72, $\eta$ = 0.71; bottom (red): $\phi$ = 1.47, $\eta$ = 0.71. $f_{\text{OSC}}$: primary oscillation frequency; $f_{\text{FI}}$: flickering instability frequency. (b) Total harmonic power analysis for the three pulsating flames.}
    \label{Fig_03}
\end{figure*} 

Some representative OH* NPSD spectra of pulsating flames are illustrated in Fig. \ref{Fig_03}. Depending on the specific flow rates of fuel, O$_2$, and CO$_2$, the flame oscillations can exhibit distinctly different characteristics. The top panel of the figure presents a typical example of fuel-rich, undiluted (or weakly diluted), low-speed flames, where both the heat release rate and the mean flow velocity are relatively low. The NPSD spectrum contains equally spaced harmonics of a single fundamental frequency around 84 Hz, with no other frequency content observed. These characteristics indicate that the flame oscillates in a single periodic mode, as corroborated by the single peak in the total harmonic power spectrum shown in the right panel.

At higher flow velocities, a second, low-frequency oscillation appeared near 10 Hz and produced sidebands around the primary oscillation frequency, as shown in the middle panel of the figure. Across the conditions explored in the current study, the second oscillation frequency remained relatively stable (approximately 9–12 Hz) and showed little dependence on equivalence ratio or dilution, indicating a mechanism distinct from the high-frequency thermo-diffusive pulsation. Further analysis reveals that this low-frequency mode arises from buoyancy-driven flame flicker, as discussed later in Section 3.3.

At higher heat release rates (i.e., higher fuel flow rates and/or lower equivalence ratios), the sidebands around the primary oscillation frequency of the flame further split into two or more fundamental oscillation frequencies, each with its own harmonics, as shown in the bottom panel of Fig. \ref{Fig_03}. In the current study, this phenomenon is referred to as multi-mode oscillation, with the corresponding primary oscillation frequency defined as the fundamental frequency of the stronger mode. Three representative cases of different flame oscillation modes have been selected for further analysis, to be discussed in the next section.

\subsection{Single-mode pulsating flames \label{sec:spectra}}

\begin{figure*}[h!]
    \centering
    \includegraphics[width= 0.9\linewidth]{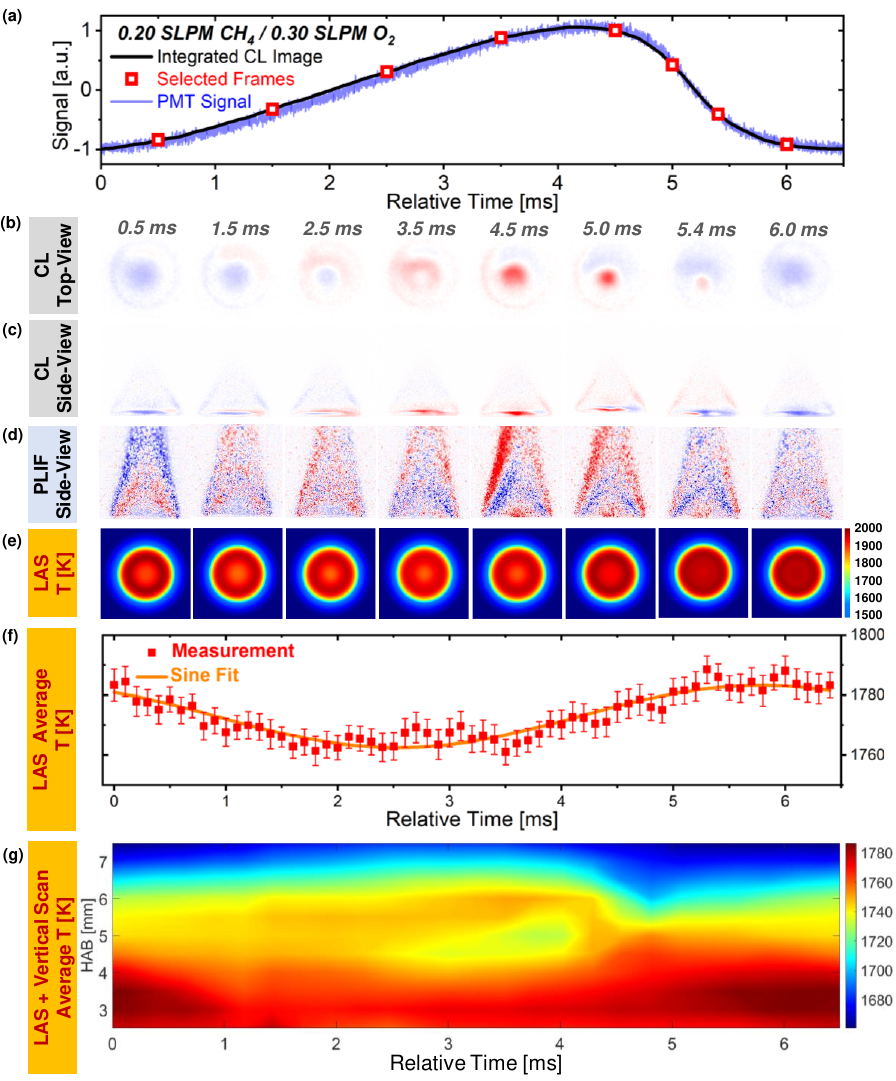}
    \caption{\footnotesize Spatiotemporally resolved measurements of a single-mode pulsating flame ($\phi$ = 1.33, $\eta$ = 1.00) at different times in its oscillation cycle. (a) Spatially integrated chemiluminescence signals from the high-speed camera (black) and the PMT (blue), normalized and phase-locked averaged over 90 cycles. (b) Phase-locked averaged top-view chemiluminescence images. (c) Phase-locked averaged side-view chemiluminescence images. (d) Instantaneous side-view OH-PLIF images. (e) Dynamic evolution of the gas temperature distribution at HAB = 2.5 mm. (f) Evolution of the average temperature along the burner diameter. Error bars denote the standard deviation of the measurements. (g) Evolution of the vertical distribution of average gas temperature along the burner diameter within a single oscillation cycle.}
    \label{Fig_04}
\end{figure*}

Fig. \ref{Fig_04} presents the dynamic evolution of multiple scalar fields in a single-mode pulsating flame under the conditions of $\dot{m}_{\rm CH_{4}}$ = 0.20 SLPM and $\dot{m}_{\rm O_{2}}$ = 0.30 SLPM. The top panel of the figure illustrates the temporal evolution of the chemiluminescence signal in a cycle, where phase-locked averaging is applied to reduce random fluctuations and improve the signal-to-noise ratio. The data are normalized to a relative intensity scale of -1 to 1. Excellent agreement is observed between the pixel-integrated signal of chemiluminescence image and the PMT signal. Within each oscillation cycle (approximately 65 ms), the total chemiluminescence is seen to increase during the first 45 ms and decrease during the last 20 ms.

Eight representative frames of the cycle-averaged chemiluminescence images are shown in Fig. \ref{Fig_04}(b) in pseudocolors, with red and blue representing high and low chemiluminescence signals, respectively. The mean intensity has been subtracted from each pixel to enhance contrast and highlight dynamic changes. From these images, the pulsating flame is seen to oscillate in a propagating ring mode, with concentric waves of local heat release rate perturbation emanating from the edge and traveling toward the flame center, where they merge and terminate before the next oscillation cycle.

Although the chemiluminescence images are line-of-sight integrated, no significant cancellation effect is expected in the top-view images because the flame front remains effectively a thin sheet throughout the oscillation cycle. This is corroborated by side-view measurements of a single-mode pulsating flame, as shown in Fig. \ref{Fig_04}(c).

OH-PLIF images in Fig. \ref{Fig_04}(d) provide additional detail on the flame structure downstream of the flame front, which is largely obscured in chemiluminescence imaging by rapid quenching of OH* to OH. From the OH-PLIF measurements, alternating regions of higher and lower OH signal are observed to move horizontally and vertically within an oscillation cycle, suggesting a traveling-wave flame oscillation mode. Unlike one-dimensional freely propagating flames, a circular porous-plug burner can support two-dimensional spatiotemporal modes, as suggested by previous studies \cite{pearlman1997target, li2025two}. This is further corroborated by flame temperature measurements in the present work.

Figure \ref{Fig_04}(e) shows the evolution of the temperature distribution at a height above the burner (HAB) of 2.5 mm, where a ring-type oscillation is evident. Under the experimental conditions of this study, the maximum local temperature is approximately 2000 K, significantly lower than the adiabatic flame temperature ($\approx$ 3000 K) due to heat loss at the water-cooled burner surface.

Fig. \ref{Fig_04}(f) shows the evolution of the line-of-sight-averaged temperature within one oscillation cycle. The measurements achieve a high level of precision (on the order of ±5 K) through cycle averaging, enabling detection of small variations. The average temperature oscillates between 1760 and 1790 K, with a phase delay of 100 - 150 degrees relative to the chemiluminescence signal. The overall 1-$\sigma$ uncertainties in these measurements are estimated to be 2 \%, based on contributions from the spectral fitting ($\leq$ 0.1\%), spectroscopy model ($\leq$ 1\%), and small fluctuations in the laser intensity baseline ($\leq$ 1\%).

\begin{figure*}[!h]
    \centering
    \includegraphics[width= 0.9\linewidth]{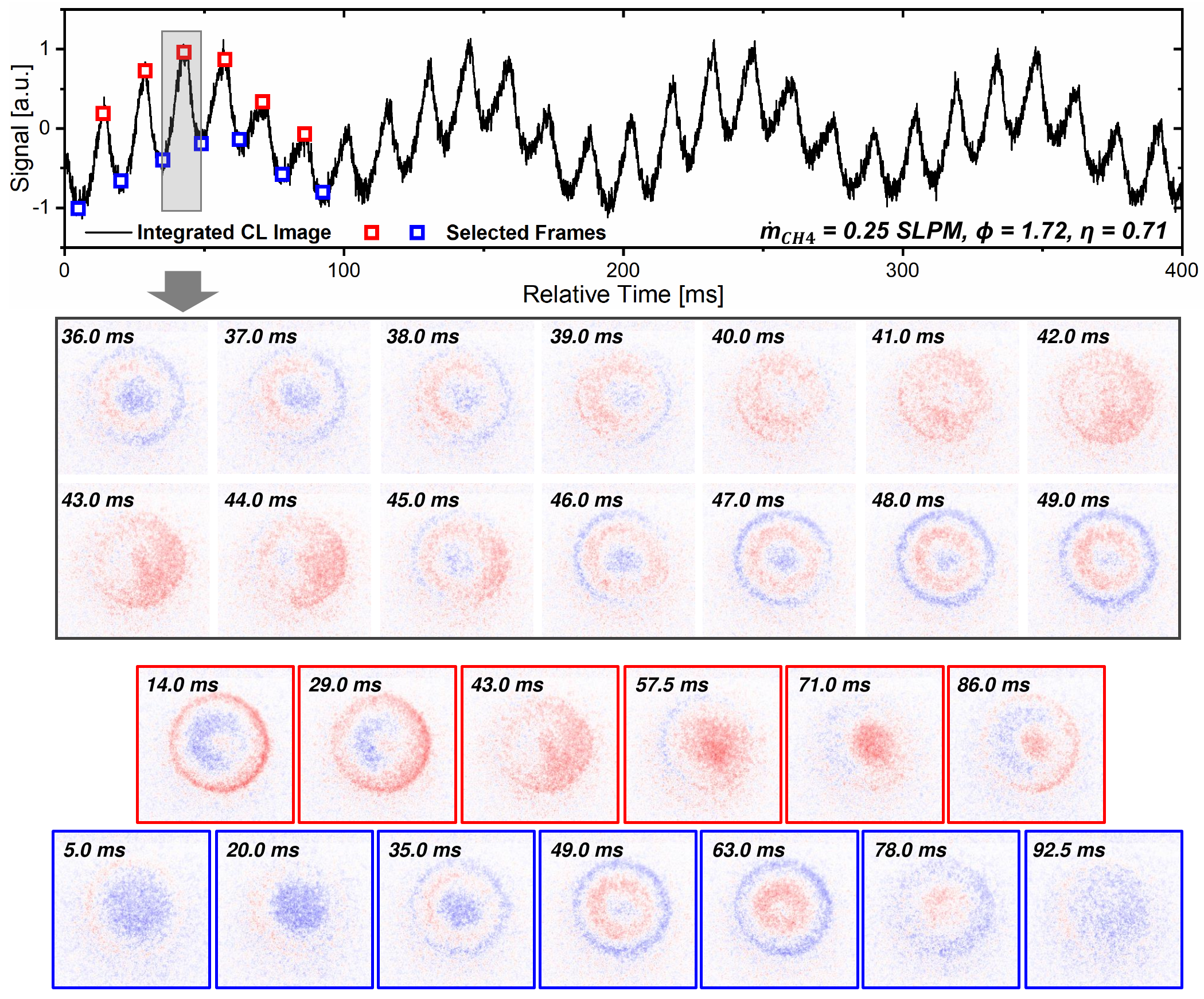}
    \caption{Chemiluminescence signal illustrating the spatiotemporal structure of a single-mode pulsating flame influenced by flickering instabilities at $\phi$ = 1.72, $\eta$ = 0.71. Selected frames of the cycle-averaged chemiluminescence images, after subtracting the mean intensity, are shown in pseudocolors, with red representing high signals and blue representing low signals.}
    \label{Fig_05}
\end{figure*}

The vertical distribution of the average temperature along the burner diameter, measured at different heights above the burner (HAB) and at various times during a single oscillation cycle, is shown in Fig. \ref{Fig_04}(g). The presence of alternating higher and lower temperature regions and their vertical motion over time correlates with the PLIF signal in Fig. \ref{Fig_04}(d), indicating traveling puffs of hotter and cooler gases in the vertical direction with a phase delay similar to that observed in Fig. \ref{Fig_04}(f). This phase delay suggests that the flame oscillation is not driven by heat loss to the burner surface alone, and an additional mechanism -- such as differential mass and heat transport -- likely interacts with heat loss to sustain the oscillation cycle. Further investigation into the 2D/3D structure of the traveling flame fronts and the detailed mechanism of flame oscillation, is warranted for future studies.

\subsection{Influence of buoyancy-driven flame flickering}
Fig. \ref{Fig_05} shows a representative case of pulsating flames with more complex dynamics. In this case, the dominant flame oscillation mode remains a ring-shaped traveling wave, while a second, low-frequency oscillation introduces a slow modulation of the primary pulsation. 

\begin{figure}[h!]
    \centering
    \includegraphics[width=\linewidth]{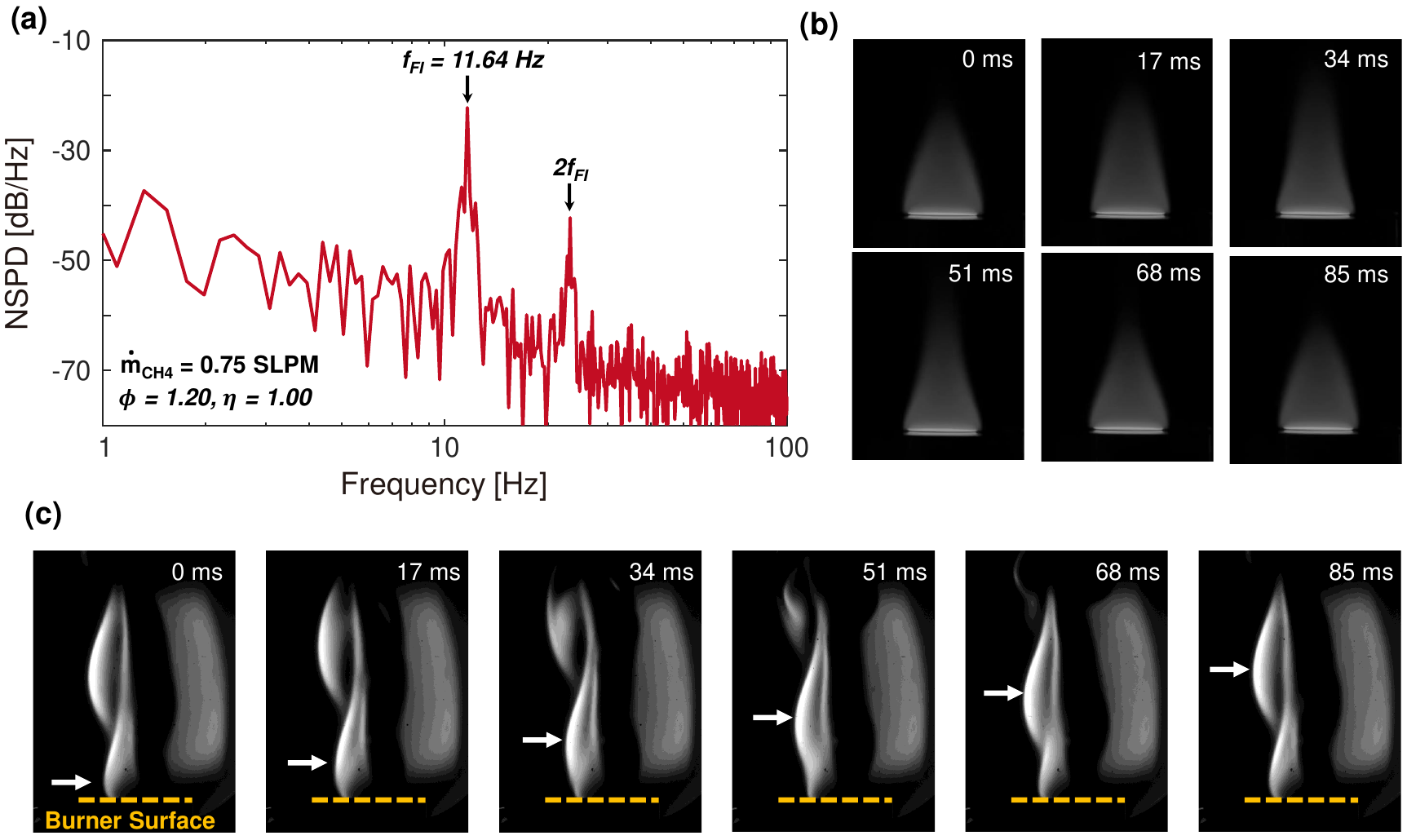}
    \caption{Example low-frequency oscillation of a methane-oxygen flame at $\phi$ = 1.20, $\eta$ = 1.00. (a) FFT spectrum of OH* normalized power spectral density (NPSD). (b) Sequential side-view images of the flame. (c) Side-view schlieren images showing toroidal vortex formation and shredding.}
    \label{Fig_06}
\end{figure}

Additional measurements of this low-frequency oscillation were conducted for an undiluted, moderately fuel-rich flame with equivalence ratio $\phi$ = 1.20, as shown in Fig. \ref{Fig_06}. Under these conditions, no high-frequency pulsation was observed; instead, a single low-frequency oscillation at a fundamental frequency of 11 Hz was detected. The results indicate that the low-frequency component corresponds to flickering-type flame dynamics, distinct from flame-front pulsation. For example, during a flame flickering cycle, the base of the flame remained largely stationary, while the tail of hot burnt gas downstream of the flame exhibited an oscillatory elongation and contraction along the axial direction.

These observed low-frequency oscillations are consistent with the general frequency range reported for buoyancy-controlled flickering of diffusion flames in early work by Buckmaster and Peters \cite{buckmaster1988infinite}. The overall dynamics align qualitatively with the planar-visualization study of Chen et al. \cite{chen1989buoyant}, in which flickering instability correlates with the periodic shredding of large toroidal vortices generated by buoyancy-driven Kelvin-Helmholtz instability outside the luminous flame. The vortex-dynamical perspective on flame flickering instability has been revisited by numerous studies across different fuels, equivalence ratios, burner diameters, and gravity levels (for example, \cite{sato2000diffusion, sahu2009effects, xia2018vortex, Yang2023}). A consensus has emerged that flickering is primarily driven by buoyant flow instability and toroidal vortices, at least at sufficiently small Reynolds numbers, and is distinct from flame extinction and re-ignition or direct modulation of the flame’s chemical reaction/heat release rate. In this sense, flickering instability is not restricted to diffusion flames but applies to premixed flames as well. The present work thus refers to this low-frequency oscillation as flame flickering instability, with its frequency denoted as $f_{\rm FI}$.

\begin{figure}[h!]
    \centering
    \includegraphics[width= \linewidth]{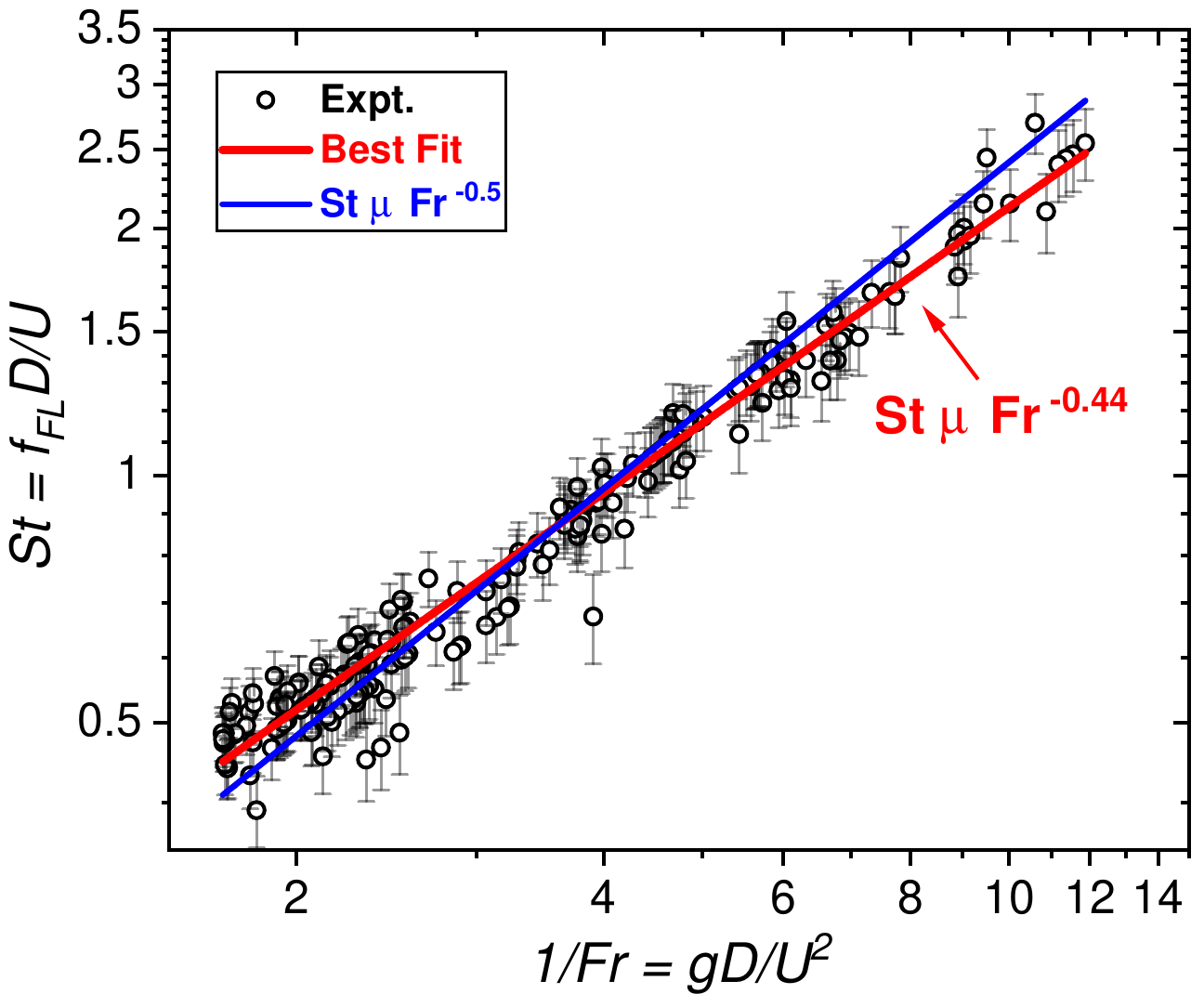}
    \caption{\footnotesize Scaling of the measured non-dimensional flame flickering frequency, $St=f_{\mathrm{FI}}D/U$, with the inverse Froude number, $1/Fr=gD/U^2$, for flames exhibiting flickering only, without thermo-diffusive pulsations. The red line represents the best fit to the measurement data, $St\propto Fr^{-0.44}$, while the blue line represents the reference scaling, $St\propto Fr^{-0.5}$.}
    \label{Fig_07}
\end{figure}

Previous studies on the flickering frequency of buoyant diffusion flames (for example, \cite{trefethen1990some, bejan1991predicting, hamins1992experimental, cetegen1993experiments, malalasekera1996review, joulain1998behavior}) have suggested the Strouhal number ($St =f_{FI}D/U$, where $D$ is the burner diameter and $U$ is the unburnt gas velocity) correlates with the Froude number ($Fr = gD/U^2$) as $St \propto Fr^{-0.5}$, in the asymptotic limit of $Fr$ approaching zero. In the present study, a total of 224 flame experiments were conducted for flickering cases without thermo-diffusive pulsation. The non-dimensional frequency, expressed in terms of $St$, is plotted as a function of $1/Fr$ in Fig. \ref{Fig_07}. Within the experimental uncertainty of the current measurements, the best-fit curve to the current data agrees reasonably well with the classic scaling of buoyancy-controlled flickering.

The present study also examined several other mechanisms that could potentially drive the low-frequency flame oscillations, including: (a) large-scale hydrodynamic shedding, (b) acoustic coupling, and (c) upstream forcing. However, their effects were found to be negligible. For hydrodynamic vortex shedding, the observed Strouhal number ($St$) varies significantly across different flame conditions, which is markedly different from the classical, nearly constant $St$ associated with vortex shedding from bluff bodies. Regarding acoustic coupling, the resonant acoustic frequency of the burner, estimated from the speed of sound and the effective cavity length inside the burner, is approximately 1250 Hz -- two orders of magnitude higher than the observed flickering frequency. Upstream forcing was minimized by precisely controlling and continuously monitoring the mass flow rates throughout the experiments, and no upstream oscillation near 10 Hz was observed.

\subsection{Multi-mode pulsating flames \label{sec:spectra}}

\begin{figure*}[!h]
    \centering
    \includegraphics[width=0.9\linewidth]{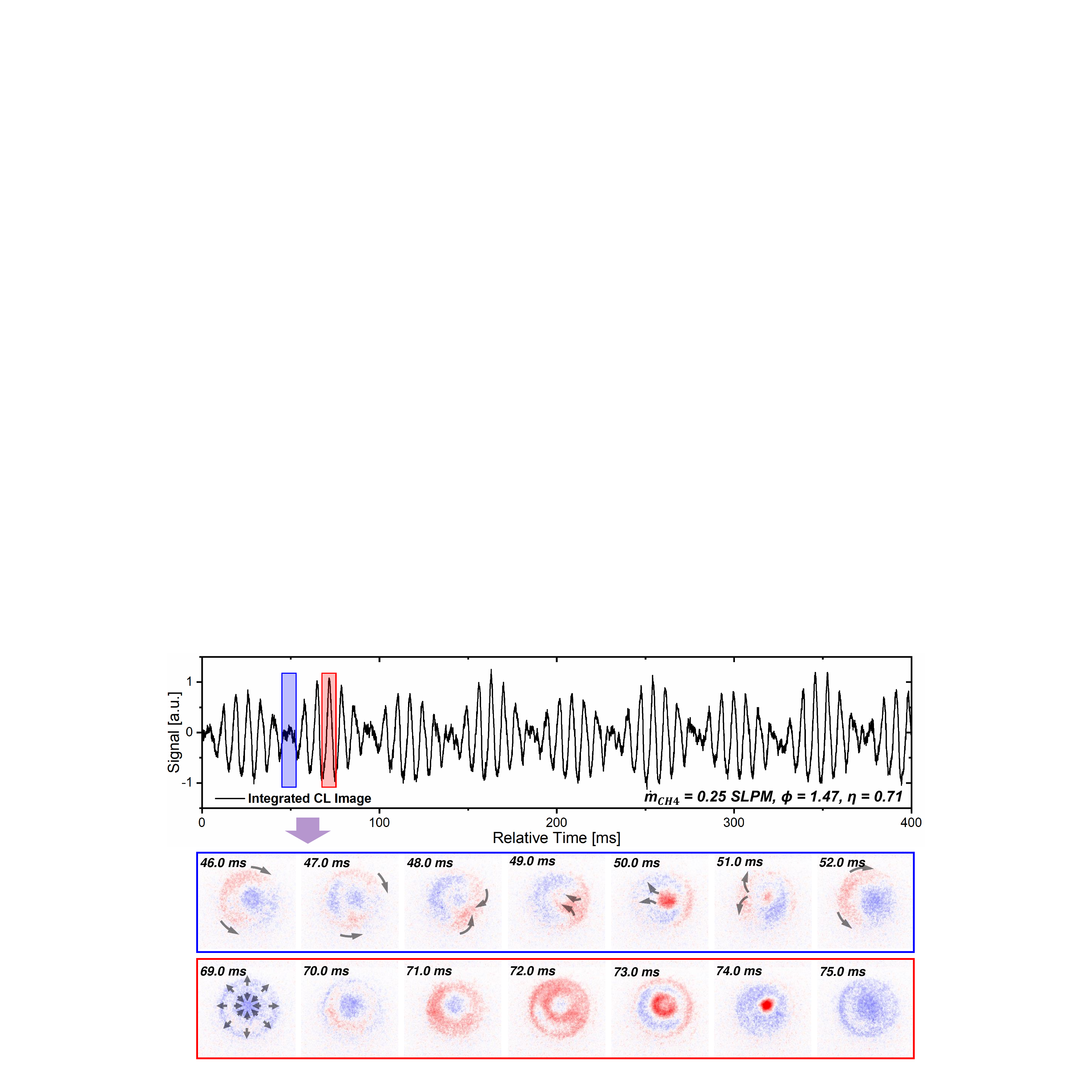}
    \caption{Chemiluminescence signal illustrating the spatiotemporal structure of a multi-mode pulsating flame at $\phi$ = 1.47, $\eta$ = 0.71.}
    \label{Fig_08}
\end{figure*}

Fig. \ref{Fig_08} shows a representative case of multi-mode flame pulsation that qualitatively differs from Fig. \ref{Fig_05}. The integrated chemiluminescence signal exhibits a clear beat pattern, indicating the coexistence of two fundamental oscillation frequencies. During constructive interference, the oscillation pattern is approximately axisymmetric, with perturbation waves emanating from a ring zone and propagating toward both the center and the burner edge. During destructive interference, the spatial pattern changes markedly: perturbation waves travel along two separate loops near the upper and lower edges of the burner and converge along the diameter. These observations suggest that the multi-mode state is consistent with a modulation-induced nonlinear interaction, in which the sidebands generated by low-frequency flickering evolve into separate pulsation modes with altered spatial structure, as the flow rate and thermal power increase. Although the frequency difference between these pulsation modes appears to be locked by harmonics of the flickering frequency (as shown in Fig. \ref{Fig_03}(c), where the difference between $f_{osc,1}$ and $f_{osc,2}$ is close to 2$f_{\rm FI}$), each mode has a separate branch of harmonics, a clear indication of mode splitting. More complex interactions among three or more modes have also appeared at conditions of higher heat release rates.

\begin{figure}
\centering
\includegraphics[width=\linewidth]{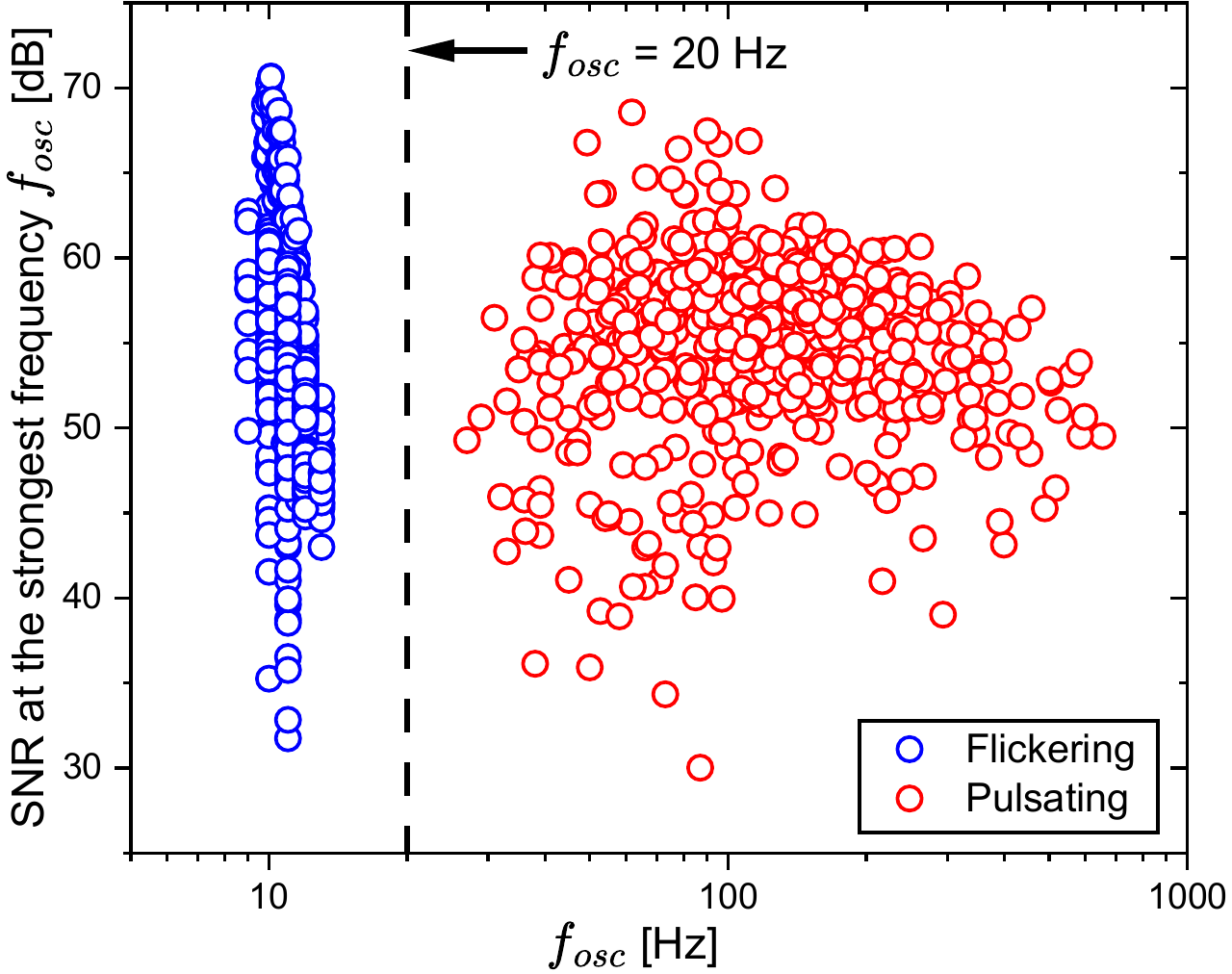}
\caption{\footnotesize Joint distribution of the observed $f_{osc}$ and SNR from the current experiments.}
\label{Fig_09}
\end{figure}

In the present study, a flame was classified as stable, flickering, single-mode, or multi-mode pulsating based on three parameters: (1) the dominant/strongest frequency, $f_{osc}$, defined by Eqn.\ref{Eqn3}; (2) the signal-to-noise ratio (SNR) at $f_{osc}$, evaluated in relative units of dB with respect to the average noise floor ($\hat{S}_{noise}$ = -80 dB/Hz) across 100 - 500,000 Hz, i.e., SNR $= \hat{S}(f_{osc})-\hat{S}_{noise}$; and (3) the relative spectral power ($R$) of the second-strongest pulsating frequency ($f_{osc,2}$) with respect to the strongest frequency ($f_{osc}$), i.e., $R = \hat{S}(f_{osc,2})-\hat{S}(f_{osc})$. First, a flame was classified as oscillating if SNR $>$ 30 dB; otherwise, it was considered stable. Note that the main conclusions were not sensitive to the specific choice of this threshold, as the qualitative results remained largely unchanged when its value was varied within $\pm$ 10 dB. Then, an oscillation was classified as flickering if $f_{osc} <$ 20 Hz, as illustrated in Fig. \ref{Fig_09}; otherwise, it was classified as pulsating. Finally, a pulsating flame was classified as multi-mode if $R > $ -10 dB (i.e., the spectral power of the second strongest mode is higher than 10\% of the strongest mode); otherwise, it was classified as single-mode (if no threshold is preferred, $R$ itself provides an objective metric of mode purity).

Dynamic Mode Decomposition (DMD) analysis was also applied to selected OH* image sequences, which yielded primary frequencies that agree with the previous results (within +/-1 Hz; see Supplementary Material II for further details). Note, however, that the DMD analysis should be regarded as a mathematical tool that seeks a best-fit approximation to the measurement data but does not always guarantee physically meaningful results, especially in the presence of experimental uncertainties. By contrast, the current study primarily relied on the phase-locked method for cycle-averaged analysis.

\begin{figure*}[h!]
\centering
\includegraphics[width=0.9\linewidth]{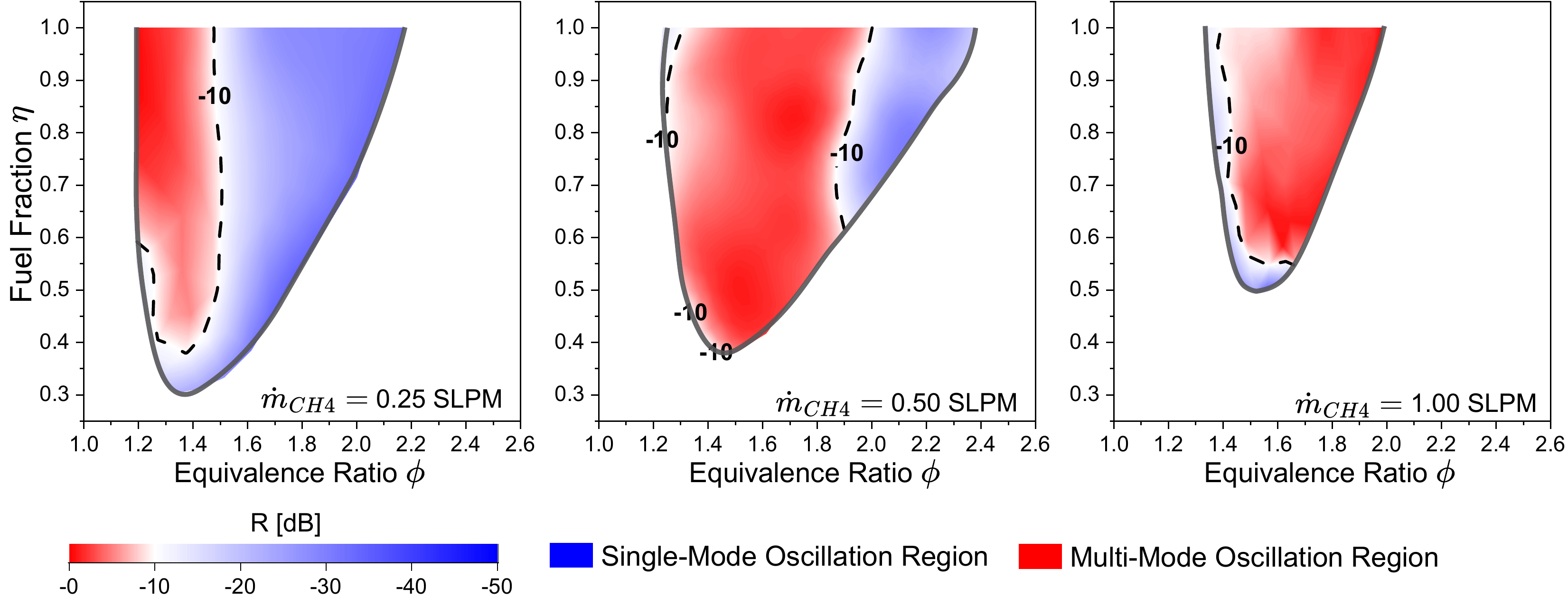}
\caption{\footnotesize Regime diagrams of pulsating instability modes at selected CH$_4$ mass flow rates. Solid lines: the observed stability boundaries. Dash lines: the boundary between single-mode and multi-mode oscillations, defined by $R = -10$ dB.}
\label{Fig_10}
\end{figure*}

\begin{figure*}[h!]
    \centering
    \includegraphics[width=0.9\linewidth]{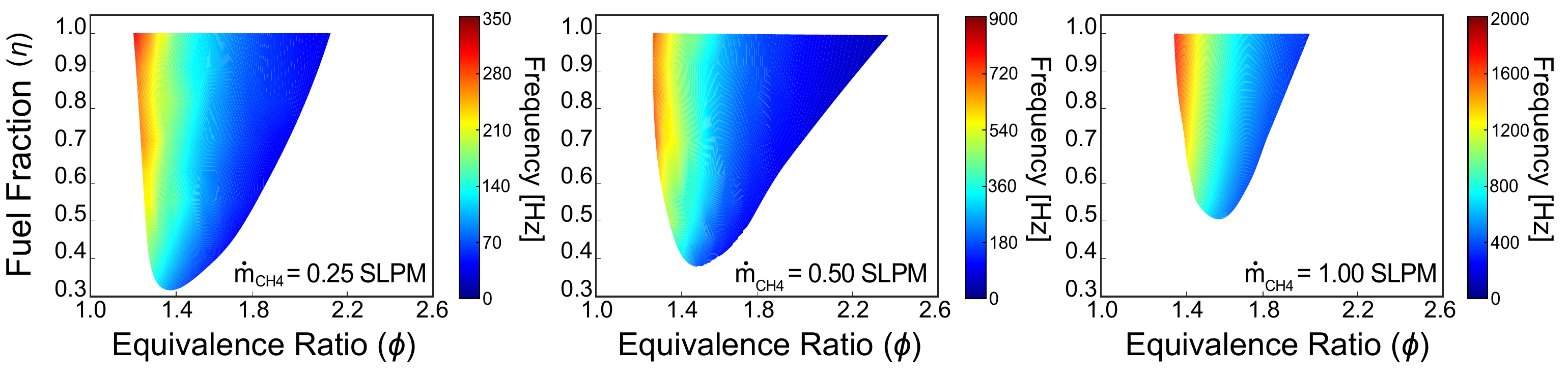}
    \caption{Parametric dependence of the primary oscillation frequency of a pulsating CH$_4$-O$_2$-CO$_2$ flame.}
    \label{Fig_11}
\end{figure*}

\subsection{Regime diagrams \label{sec:spectra}}
The primary frequencies and modes of pulsating CH$_4$-O$_2$-CO$_2$ flames, measured over a range of fuel flow rates, equivalence ratios, and fuel fractions, are provided in Supplementary Material I. The regime diagrams of pulsation modes and frequency contour plots for three representative fuel flow rates are displayed in Figs. \ref{Fig_10} and \ref{Fig_11}. Under the present experimental configuration, pulsating flame instability occurred only for fuel-rich conditions, within the equivalence ratio range of 1.18 to 2.50. The range of unstable equivalence ratios narrows with increasing CO$_2$ dilution (i.e., decreasing $\eta$), while its dependence on the CH$_4$ flow rate is non-monotonic: it first broadens with increasing CH$_4$ flow rate, peaks around 0.4 SLPM (see Supplementary II for additional plots), and then contracts to a single point. This trend is consistent with results reported in \cite{kurdyumov2008porous}. The highest CH$_4$ flow rate at which flame pulsation was observed is approximately 1.32 SLPM, corresponding to $\phi$ = 1.48 and $\eta$ = 1.00.

The single-mode oscillation regimes are generally narrower than the multi-mode regimes, except at low CH$_4$ flow rates ($\leq$ 0.25 SLPM), where heat losses to the burner are significant. At CH$_4$ flow rates of 0.5 SLPM or higher, the single-mode regimes are confined to regions near the stability boundaries. The primary oscillation frequency tends to increase with the mass flow rate of CH$_4$ and decrease with the equivalence ratio, showing a strong correlation with the flame's thermal power. By contrast, the influence of CO$_2$ dilution on the primary oscillation frequency is less significant. It is noteworthy that, although heat loss modulates the burned-gas temperature and the ranges of equivalence ratio and fuel fraction for thermo-diffusive pulsations, it has little effect on the frequency of buoyancy-induced flame flicker ($\sim$ 10 Hz).

Compared with freely propagating flames where thermo-diffusive pulsations occur only at high Lewis numbers and Zel’dovich numbers, in the present study they are observed over a much wider range of conditions. This observation suggests that heat loss associated with the cool burner surface significantly promotes thermo-diffusive pulsations. In the present study, $Le_{\mathrm{eff}}$ is defined as an overall effective Lewis number following the excess/deficient-reactant formulation proposed in earlier asymptotic and flame-stability analyses (e.g., \cite{bechtold2001dependence, addabbo2002wrinkling}) and in the recent study by Wang et al. \cite{ wang2023revisiting}:
\begin{equation} \label{Eqn4}
Le_{\mathrm{eff}}=1+\frac{\left(Le_{\mathrm{exc}}-1\right)+\beta\left(Le_{\mathrm{def}}-1\right)}{1+\beta},
\end{equation}
where $\beta=1+Ze(\Phi-1)$, $\Phi$ equals $\phi$ (or $1/\phi$) for fuel-rich (or fuel-lean) mixtures,  $Le_{\mathrm{exc}}$ and $Le_{\mathrm{def}}$ denote the Lewis numbers of the excess and deficient reactants, respectively. The Lewis number of each reactant is evaluated from the unburned-gas thermal diffusivity and the mixture-averaged mass diffusivity. The transport and thermochemical properties are obtained from Cantera calculations \cite{cantera} using the DRM19 mechanism \cite{DRM19}. The Zel'dovich number, defined as $Ze=E_a(T_b-T_u)/R T_b^2$, is also calculated following the method of Wang et al.~\cite{wang2023revisiting}. In this equation, $R$ is the universal gas constant, $T_u$ is the unburned-gas temperature measured experimentally with the thermocouple embedded in the porous plug, $T_b$ is the burned-gas temperature estimated from the maximum temperature of one-dimensional burner-stabilized flame calculations using Cantera, and $E_a$ is the overall activation energy, which is regarded as an intrinsic property of the reactive gas mixture and is evaluated from adiabatic flame calculations. Representative results at $\dot{m}_{CH_4}$ = 0.50 SLPM are shown in Fig. \ref{Fig_12}, and additional results can be found in the Supplementary Material II.

Under the conditions explored in the present study, the maximum value of $Ze(Le_{\mathrm{eff}}-1)$ is slightly larger than 2 and significantly less than $4(1+\sqrt{3})$ -- the critical value for adiabatic, freely propagating flames in homogeneous gas mixtures. Pulsating flames were observed at near-unity Lewis numbers, with a few high-CO$_2$-dilution cases at $Le_{\mathrm{eff}}$ slightly below unity. A plausible explanation is that the presence of the bronze porous plug modified the effective strength of differential diffusion, as it significantly promoted heat transfer without substantially affecting species diffusion. The overall effect on flame dynamics was similar to increasing $Le_{\mathrm{eff}}$, and consequently, the nominal stability boundary extended to much lower $Le_{\mathrm{eff}}$ than predicted by the classical Sivashinsky criterion.

\begin{figure}[h!]
\centering
\includegraphics[width=0.9 \linewidth]{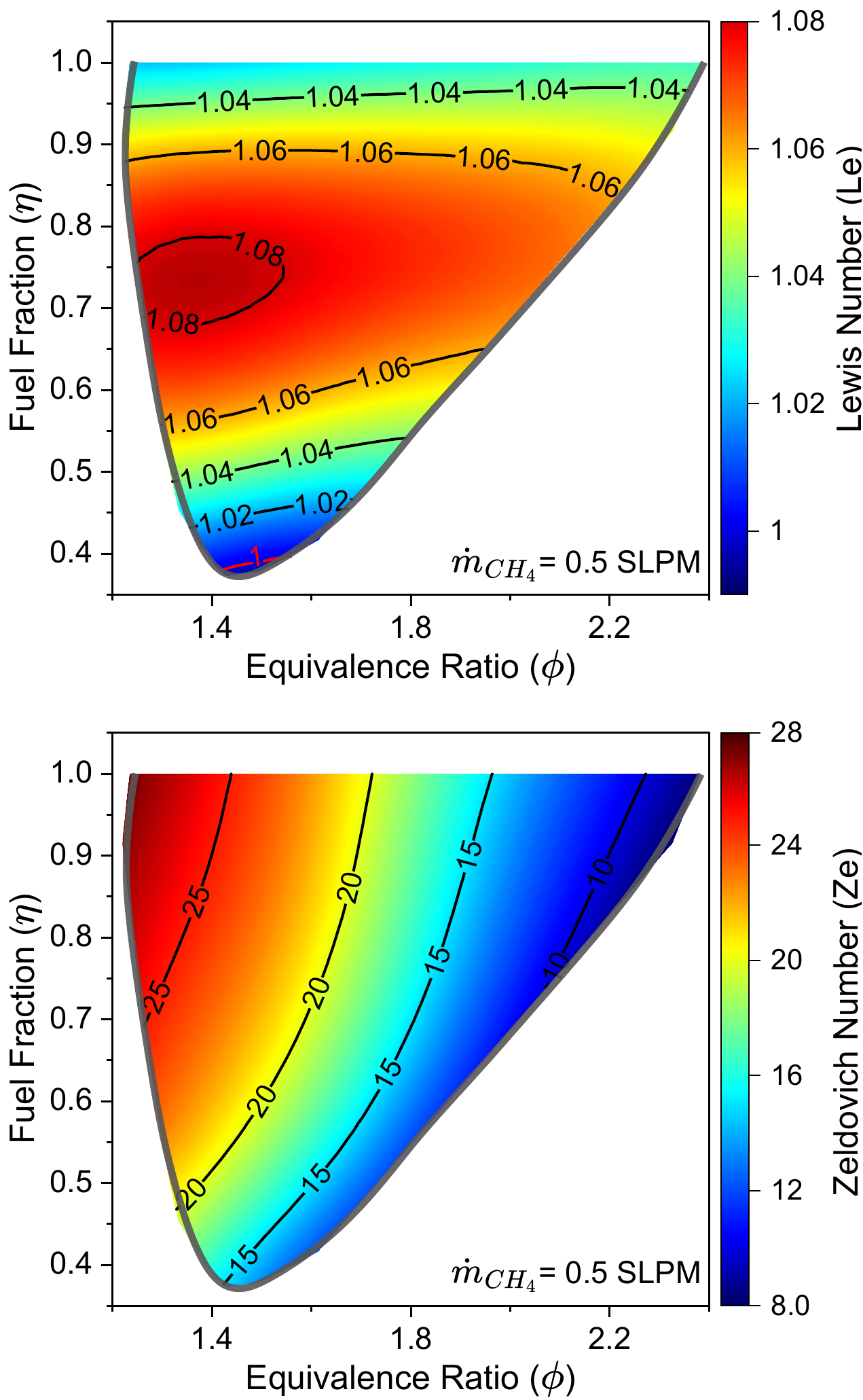}
\caption{\footnotesize Effective Lewis number ($Le_{\mathrm{eff}}$) and Zel'dovich number ($Ze$) as functions of equivalence ratio ($\phi$) and fuel fraction ($\eta$), calculated from the measured $T_u$ at $\dot{m}_{CH_4}$ = 0.50 SLPM using Cantera \cite{cantera} and the DRM19 mechanism \cite{DRM19}.}
\label{Fig_12}
\end{figure}

\section{Conclusions \label{sec:conclusions}}
The pulsating instabilities of premixed CH$_4$-O$_2$ flames on a porous-plug burner were investigated across varying levels of CO$_2$ dilution, equivalence ratios, and flow rates. Heat loss to the flat burner surface amplified the thermo-diffusive instabilities and enabled pulsations at near-unity Lewis numbers, where freely propagating flames would be stable.

In this study, multiple modes of flame pulsation were observed. Under fuel-rich conditions with relatively low heat release and flow rates, the flames exhibited quasi-periodic pulsation in a single mode. At higher flow rates, buoyancy-controlled, low-frequency flickering instability appeared and began to interfere with the thermo-diffusive pulsation, producing sidebands around the primary frequency. At elevated heat release rates, nonlinear interaction between flame flicker and single-mode pulsation further triggered mode splitting, giving rise to multi-mode pulsation. The stability boundaries and regime diagrams of flame oscillation modes were established, and representative measurements of the spatiotemporal variations in pulsating flames were analyzed. Further investigation into the transient dynamics of mode transitions is in progress, particularly under microgravity conditions to eliminate buoyancy effects.

The observed trends of stability boundary and mode distributions are expected to generalize to other CO$_2$-diluted oxy-fuel combustion systems. While this study focuses on methane flames, the underlying concepts can be extended to other fuels, such as hydrogen and hydrogen-blended fuels. In this sense, the findings of the present study are useful for both fundamental research on flame dynamics and practical applications of oxy-combustion.

\section*{CrediT authorship contribution statement}
\textbf{Xiangyu Nie:} Data Curation, Formal Analysis, Investigation, Writing - Original Draft. \textbf{Shuoxun Zhang:} Data Curation, Investigation. \textbf{Shengkai Wang:} Conceptualization, Funding acquisition, Methodology, Supervision, Writing - review and editing.

\section*{Declaration of competing interest}
The authors declare that they have no known competing financial interests or personal relationships that could have appeared to influence the work reported in this paper.

\section*{Acknowledgments}
This research is supported by the National Key Research and Development Program of China under Grant No. 2025YFF0511801 and No. 2021YFA0717200, the National Natural Science Foundation of China under Grants No. 12472278 and No. 92152108, the Space Application System of China Manned Space Program under projects, and the Open Research Program of National Key Laboratory of Fundamental Algorithms and Models for Engineering Simulation. Numerical simulations were supported by the High-Performance Computing Platform of Peking University.

\section*{Supplementary material}
Supplementary Material I: Data summary of flame oscillation frequencies and modes under conditions explored in the current study. Supplementary Material II: Representative results of dynamic mode decomposition analysis, regime diagrams, primary oscillation frequency, effective Lewis number and Zel'dovich number at additional CH$_4$ mass flow rates.

\FloatBarrier

\bibliographystyle{cnf-num}
\bibliography{References}

\end{document}